# Hierarchical cross-linking in physical alginate gels: a rheological and dynamic light scattering investigation

*Domenico Larobina\*[a,b] and Luca Cipelletti\*[c,d]*

[1] Institute of Composite and Biomedical Material – National Research Council of Italy, CNR – P.le E. Fermi 1, I-80055 Portici, Naples, Italy, and INSTM UdR of Naples, Italy

[2] Université Montpellier 2, Laboratoire Charles Coulomb UMR 5221, F-34095, Montpellier, France

[3] CNRS, Laboratoire Charles Coulomb UMR 5221, F-34095, Montpellier,

We investigate the dynamics of alginate gels, an important class of biopolymer-based viscoelastic materials, by combining mechanical tests and non-conventional, time-resolved light scattering methods. Two relaxation modes are observed upon applying a compressive or shear stress. Dynamic light scattering and diffusive wave spectroscopy measurements reveal that these modes are associated with discontinuous rearrangement events that restructure the gel network via anomalous, non-diffusive microscopic dynamics. We show that these dynamics are due to both thermal activation and internal stress stored during gelation and propose a scenario where a hierarchy of cross-links with different life times is responsible for the observed complex behavior. Measurements at various temperatures and sample ages are presented to support this scenario.

INTRODUCTION

Alginates are polysaccharide biomolecules that can form gelled structures under appropriate conditions. Alginate-based materials are attractive because of their bio-compatibility, non-toxicity and bio-degradability, which make them ideal for pharmaceutical, food, and bioengineering applications.[1,2] In many of these applications knowledge of the mechanical and structural properties of the gel plays a crucial role. For instance, in tissue engineering applications a change in the mechanical and swelling properties of the gel may result in enhanced cartilage tissue properties[3,4]; similarly, in pharmaceutics the release of drug is greatly affected by how the gel stress evolves during application.[5] In view of their relevance for many applications, the mechanical properties of alginate gels have been extensively studied.[6,7,8] By contrast, much less is known on the relaxation processes at the microscopic level, which are ultimately responsible for the viscoelastic properties of the gels.

Sodium alginate has a block copolymer structure, being formed by β-L-mannuronic acid (M) and α-L-guluronic acid (G) units that can be ionically cross-linked by divalent ions.[9] In particular, the stereochemistry of the G monomer allows for the formation of a particularly stable cross-link structure, known as "egg-box".[10] In the simplest egg-box structure four G monomers, belonging to two alginate strands, are connected by a divalent ion through coordination of four carboxylic groups. Actually, such links are mostly favorite among G blocks of more than 8 units (rod-like junctions[11,12]), and can even give rise to bundles of junctions with more than two chains involved, as the divalent ion concentration increases.[13] Even if the egg-box junctions are determinant to the formation of a solid-like gel, it is worth noticing that other ionic cross-links, e.g. between M-M and M-G units, although less stable, are equally likely.[14]

Due to this spectrum of ionic cross-links of different stability, alginate gels have mechanical properties that are determined not only by alginate concentration, but also by the ratio between G and M blocks, and the divalent ion concentration.[15,16,17] Moreover, the gelation protocol as well as temperature and pH conditions also influence the final structure of the network and its resulting mechanical properties.[18] Finally, the random ionic gelation process unavoidably induces some degree of heterogeneity on the mesoscopic structure of the gel. The resulting gel is then far from equilibrium and is expected to undergo a slow relaxation, or aging, towards its stable state.

From the continuum mechanics point of view, the above picture of alginate cross-links, where ionic bonds can undergo detaching and re-bonding, results in a viscoelastic behavior of the material as a whole. Such viscoelastic nature has been studied by many authors that reported alginate gel properties probed under compressive, tensile and shear stress.[16,19,20,21] These works describe alginate gels as complex viscoelastic solids, with an elastic behavior at small deformation and short times that becomes plastic at larger deformations and time scales. These features, together with the possibility to fine-tune the gel properties by changing its composition (polymer and cross-linking ions concentration, ratio of G and M blocks etc.) make alginate gels very versatile materials, an important feature in applications.

To fully control and exploit the properties of alginate-based materials, a thorough understanding of the microscopic mechanisms underlying their macroscopic properties would be desirable. Unfortunately, mechanical tests cannot directly probe such microscopic processes. By contrast, insight on the microscopic dynamics of the gels may be obtained by dynamic light scattering (DLS)[22], which probes the relaxation of density fluctuations on time scales ranging from microseconds to tens of seconds and length scales from a fraction of a $\mu$m to several $\mu$m. To avoid experimental difficulties arising from the non-ergodic behavior of gelled samples, measurements have been typically restricted to ergodic alginate solutions, prior to gelation.[23] While this provides valuable information on the aggregation process eventually leading to gelation, in particular by pointing out the role of thermally activated cross-link breaking[23], it cannot elucidate the relaxation dynamics of the gels themselves.

A remarkable exception is provided by the work of Secchi *et al.*[24], who recently applied advanced DLS methods that combine imaging and scattering[25] to study the kinetics of formation and the aging of alginate gels prepared by diffusion of $Ca^{2+}$ from an external reservoir brought in contact with the alginate solution via a porous membrane. A slow relaxation mode is observed, on the time scale of thousands of seconds, together with aging effects (slowing down of the dynamics) and intermittent bursts of dynamical activity. The physical picture proposed in Ref. 24 is that of a network that can slowly restructure because the cross-links are not permanent, possibly under the action of internal stress accumulated during the gelation process, in analogy to what observed in colloidal[26,27] and actin[28] gels. Unfortunately, the limited temporal resolution of the experiment by Secchi *et al.* did not allow for a full comparison with mechanical tests: while a broad spectrum of relaxation modes were observed for the latter, only one single relaxation mode was found in light scattering. Additionally, data were collected at a single scattering angle, so that the length-scale dependence of the dynamics could not be established, a crucial test to discriminate between anomalous, stress-induced dynamics and more conventional diffusive dynamics.[26] Such a test is particularly important given that in Ref. [24] the DLS relaxation mode was find to be exponential, in startling contrast with experimental reports on internal stress-driven slow dynamics, where correlation functions measured by light scattering typically decay steeper than exponentially ("compressed exponential"[26,29]). Finally, we note that internal stress is likely to be generated when a gelling front propagates in the sample, as in Ref. 24, since it is known that gelation by diffusion of divalent ions induces a macroscopic gradient in the gel structure and mechanical properties, because the gradient of ion concentration inevitably associated with diffusion entails a gradient in the density and nature of crosslinks.[30] By contrast, it is not clear whether such a mechanism would be relevant for gels that are prepared in a macroscopically homogeneous state via inner reaction methods.[15]

In this work, we address the issue of the relation between mechanical properties and microscopic dynamics in homogenously gelled alginate solutions and we investigate the physical mechanism responsible for the latter. In addition to rheological and dynamical mechanical analysis (DMA) measurements, we perform a variety of light scattering experiments, including DLS at four different scattering angles and Diffusing Wave Spectroscopy (DWS)[31] on turbid samples loaded with small amounts of tracer particles. Our light scattering experiments allow us to probe the gel dynamics on a wide range of length scales, from a few nm up to 1 $\mu$m, and point to a wide spectrum of relaxation processes. To better assess the role of thermal energy *vs*. internal stress in the gel dynamics, measurements at various temperatures are performed for both mechanical tests and light scattering experiments. Collectively, our findings suggest that alginate gels are formed by a network comprising a hierarchy of physical (non-permanent) bonds with different life times and that their microscopic dynamics is ruled by both internal stress relaxation and thermal activation.

MATERIALS AND METHODS

A medium viscosity alginate sample was supplied by FMC Biopolymers (U.K.). According to the specifications from the manufacturer, this sample has a 1% solution viscosity of 70.7 mPa s, with a weight-average molecular weight of 97 kDa and a content of guluronic acid of 63%. H-NMR analysis conducted on the same batch of alginate powder has essentially confirmed the composition of our sample: we found $F_G = 0.69$, $F_{GG} = 0.56$, $F_{GM} = 0.12$, where $F_G$, $F_{GG}$, and $F_{GM}$ represent the fractions of monomer G, dimers G-G, and dimers M-G, respectively. Calcium chloride, ethylene-glycol-tetraacetic acid disodium salt (Na$_2$-EGTA), and D-glucono-δ-lactone (GDL) were all purchased from Sigma and used as received.

Alginate gelation is realized by an inner reaction technique, which ensures homogeneous gelation on a macroscopic scale. Details of the preparation can be found in the literature.[15] Briefly, we start from a 0.05 molar Ca-EGTA solution prepared by adding 0.07 g (6.3·10$^{-4}$ moles) of CaCl$_2$ to 12.6 ml of Na$_2$-EGTA (0.05 M). The pH of the Ca-EGTA solution is then adjusted to 7 by adding approximately 2 ml of NaOH (1 M). Next, 0.8 grams of Alginate are dissolved in the Ca-EGTA solution, along with a small amount of the final water mass to avoid alginate clumping. When the alginate is completely dissolved, we add the rest of the water (up to 61.53 g), in order to obtain a final alginate concentration of 1.3% w/w. The gelation is induced by adding 0.23 g (about twice the moles of CaCl$_2$) of the slowly hydrolyzing GDL. This allows for thorough mixing of all components before any significant aggregation starts, thereby insuring the homogeneity of the gels. For particle-loaded gels, polystyrene latex spheres are added to the alginate solution before adding GDL. The particle diameter and concentration are $2a = 505$ nm, $\varphi_{DLS} = 0.0032\%$ and $2a = 910$ nm, $\varphi_{DWS} = 0.8\%$ for DLS and DWS, respectively.

After mixing the solution is immediately poured into: *i*) a glass cylinder of 8 mm inner diameter for DLS; *ii*) a home-made rectangular cell with thickness 5 mm or 10 mm for DWS; *iii*) a Petri dish for mechanical tests. In the latter case, the Petri dishes are exposed for 24 hours to a water saturated environment at ambient temperature before testing. During this period gelation is completed within 4-5 hours, as checked by rheological measurements. Cylindrical samples of about 5 mm in thickness are retrieved for DMA and rheological measurements as explained below.

Dynamic light scattering

Dynamic light scattering (DLS) is a popular technique widely used to probe the microscopic dynamics of soft materials such as colloidal suspensions, emulsions, surfactant phases, polymer solutions and particle or polymer gels.[22] While DLS deals with almost transparent samples for which only single scattering is relevant, Diffusing Wave Spectroscopy (DWS) has extended the method to very turbid samples, in the strong multiple scattering regime.[31] The gels studied here are nearly transparent: scattering experiments on bare gels thus probe the DLS regime. We also perform experiments on gels loaded with small amounts of tracer particles, at a volume fraction $\varphi$. We use two values of $\varphi$: for $\varphi = \varphi_{DLS} = 0.0032\%$ the scattering from the particles dominates over that from the gel, but experiments are still in the single scattering regime and the DLS formalism can be applied. By contrast, for $\varphi = \varphi_{DWS} = 0.8\%$ multiple scattering from the particles is rampant, so that data are collected in the DWS regime.

In both DLS and DWS one illuminates the sample with a laser beam and measures the autocorrelation function of the temporal fluctuations of the scattered intensity $I(t)$, $g_2(\tau) - 1 = \frac{\langle I(t)I(t+\tau)\rangle}{\langle I(t)\rangle^2} - 1$, where $\langle \cdots \rangle$ denotes an ensemble average, usually replaced by a time average in experiments. The interpretation of $g_2-1$ differs for DLS and DWS. In DLS, data are collected at a well-defined scattering vector $q = 4\pi n \lambda_0^{-1} \sin(\theta/2)$, where $n$ is the refractive index of the solvent, $\lambda_0 = 532.5$ nm the in-vacuo laser wavelength, and $\theta$ the scattering angle. This allows the dynamics to be probed on a length scale of the order of $2\pi q^{-1}$. More precisely, $g_2$- 1 is proportional to the squared intermediate scattering function that quantifies the temporal decay of density fluctuations of wave vector $q$.[22] Diffusing Wave Spectroscopy experiments, by contrast, are performed either in the transmission or backscattering ($\theta \approx 180°$) geometry[31], the exact value of $\theta$ being irrelevant. In the backscattering geometry that we use in this work and in the limit of sufficiently thick samples, the intensity correlation function measured by DWS depends on the mean squared displacement (msd) of the tracer particles, $\langle \Delta r^2(\tau) \rangle$, via an expression that is very well approximated by:

$$g_2(\tau) - 1 = \beta \exp\left[-2\gamma \sqrt{k_0^2 \langle \Delta r^2(\tau) \rangle}\right], \qquad (1)$$

where $k_0 = 2\pi \lambda_0^{-1}$, while $\gamma$ and $\beta \leq 1$ depend on the polarization status of the incident and detected light, and on the collection optics, respectively.[31] In our DWS experiments, the laser beam is linearly polarized and a polarizer is used to accept only scattered light with polarization orthogonal to that of the incoming beam, leading to $\gamma = 2$.

It should be noted that DLS and DWS are complementary methods in that they probe different length scales. In our DLS measurements, we use the setup described in Ref. 32 and choose $\theta$ = 22.5, 45, 90 or 120 deg, corresponding to $q$ = 6.12, 12.0, 22.2, and 27.2 $\mu$m$^{-1}$ and a probed length scale $2\pi q^{-1}$ = 1.03, 0.52, 0.28, and 0.23 $\mu$m, respectively. By contrast, the DWS data cover rms displacements on much smaller length scales, from a few nm up to about 1 $\mu$m.

Special care must be taken when dealing with samples that are non-ergodic, i.e. samples that explore only a limited portion of the phase space on experimentally accessible time scales. This is indeed the case of our gels, since the alginate molecules are trapped in a network that, as we shall show it, remodels itself only on time scales exceeding several hours. For such samples, the time average usually applied to measure $g_2-1$ is practically unfeasible. To circumvent this difficulty, we use for both DLS and DWS measurements the Time Resolved Correlation (TRC) method.[33] A multi-pixel CCD camera is used as a detector, instead of a single detector such as a phototube, as in traditional light scattering. The signal of each pixel is treated independently and the intensity correlation function is averaged over pixels, rather than over time. More specifically, one calculates the degree of correlation $c_I$ between two images of the scattered light taken at times $t$ and $t+\tau$:

$$c_I(t,\tau) = \frac{\langle I_p(t) I_p(t+\tau) \rangle_p}{\langle I_p(t) \rangle_p \langle I_p(t+\tau) \rangle_p} - 1, \quad (2)$$

where $I_p(t)$ is the intensity at time $t$ for the $p$-th pixel and $\langle \cdots \rangle_p$ denotes an average over CCD pixels. For stationary dynamics, $c_I$ may be further averaged over $t$ to improve statistics, yielding the usual $g_2 - 1$. If the dynamics evolve with sample age $t_w$, as for our gels, $c_I$ is averaged over a short time window $t_{ave}$ in order to obtain an age-dependent intensity correlation function:

$$g_2(t_w, \tau) - 1 = \langle c_I(t, \tau) \rangle_{t_w \leq t \leq t_w + t_{ave}} \quad (3)$$

where we typically use $t_{ave} \leq t_w / 10$. Note that the use of a CCD detector limits the smallest accessible delay to the inverse frame rate, typically of the order of several milliseconds at least. However, this is not a severe limitation for systems that exhibit very slow relaxations, as the gels studied here.

As discussed in Ref. 33, insight on the nature of the dynamics can be gained by inspecting the degree of correlation $c_I$ for a fixed time delay $\tau$, as a function of $t$. For systems at equilibrium, the degree of correlation is found to be constant, within experimental uncertainty, when plotted as a function of $t$. This reflects time translational invariance for equilibrium dynamics: the loss of correlation over a time lag $\tau$ does not depend on the starting time. By contrast, the dynamics of many jammed or glassy soft materials has been shown to be highly heterogeneous in time, with sudden drops of $c_I$ corresponding to intermittent rearrangements.[25,27,34] Thus, TRC has the capability of discriminating between continuous, equilibrium dynamics and heterogeneous, intermittent rearrangements, a feature that we will exploit in the following.

Mechanical tests

Stress relaxation properties of alginate gel are measured in compression and shear tests with a dynamic mechanical analyzer (Q 800, TA Instruments) and a rotational rheometer (AR 500, TA Instrument), respectively. Cylindrical samples with diameters of 12 mm for compression and 25 mm for shear are cut from a large gel cake of about 5 mm thickness and tested without further alteration. In the case of compressive measurements, lubricated impermeable plates are used to allow the disk to expand freely under compression; in this condition no bulging of the gel lateral surface is observed. For shear tests, serrated parallel plates are adopted to prevent slip effects. To limit sample dehydration during the measurements, wet sponges are positioned around the gel, for both DMA and shear rheology. Preliminary stress-strain (at 0.2 min$^{-1}$ strain rate) and strain-sweep (at 1 Hz) measurements are conducted in compressive and shear configuration in order to evaluate the linear regime of the material. These tests show that the stress-strain relation is linear up to a strain of 0.1 in compression, while the shear modulus in strain-sweep tests is observed to be constant only up to a strain of 0.02. Based on the results of these tests, we perform stress relaxation tests at a strain of 0.05 and 0.01 for compression and shear, respectively.

RESULTS AND DISCUSSION

Compressive stress relaxation measurements on alginate gels are reported in Fig. 1 as the Young modulus normalized with respect to its value at time $\tau=0^+$ *vs.* time, for three different temperatures (20, 37, and 50°C). Although care was taken to avoid dehydration of the sample during the tests, as discussed above, only stress values up to 2000 s are considered here.

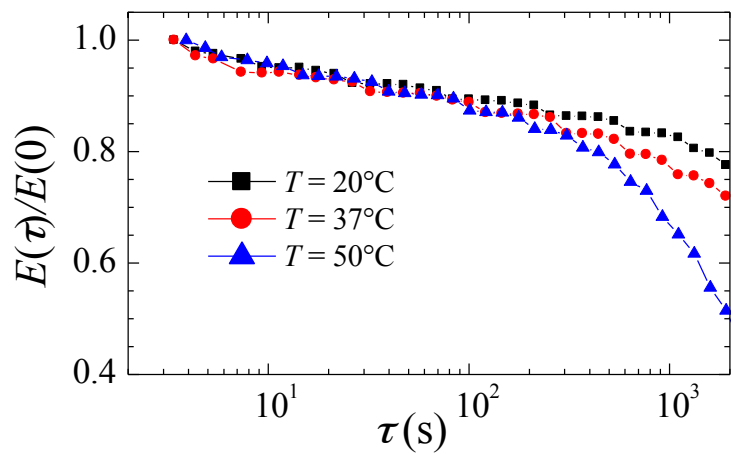

**Figure 1**. Relaxation of Young modulus after applying a compressive stress in the linear regime, for a gel with 1.3%w/w of alginate, at the three temperatures indicated by the labels. Two main relaxation modes are visible, in the range 10-100 s and for $\tau \sim 1000$ s, respectively.

The data clearly display a double decay: a first decay, occurring between 10 and 100 seconds, is followed by a second decay on time scales of 1000 s or longer. Only the initial stage of the latter is seen within the duration of the measurement. We will refer to these decays as the "intermediate" and the "final" relaxation, respectively. This terminology is justified by the analogies with light scattering measurements that will be discussed in the following. Hints of a fast mode, with a characteristic time of the order of 1 s are also visible, in particular for the data at $T = 37°C$. It is worth noticing that the intermediate decay accounts only for a modest relaxation of the applied stress, of order 10%. In other words, rearrangements associated with the intermediate relaxation mechanism are not able to fully absorb the applied deformation. Thus, some longer-lived stress-bearing structure is holding the remaining 90% of the stress, up to time scales of thousands of seconds.

Stress relaxation data in shear experiments confirm the behavior observed in compression. In analogy to compression tests, the time-dependent shear modulus $G(\tau)$ after a step strain relaxes via a two-step decay (data not shown). The "initial" modulus, estimated as $G' = G(\tau=3s)$, has a value of $0.73 \pm 0.1$ kPa, roughly a third of the initial value obtained in compression, $E' = E(\tau=3s) = 1.9 \pm 0.2$ kPa. This indicates that on very short time scales the gel, as a whole, behaves essentially as an incompressible material ($E' = 3G'$), consistently with the notion that water cannot be expelled from the gel network on short time scales, as it would be required for a compressible material.

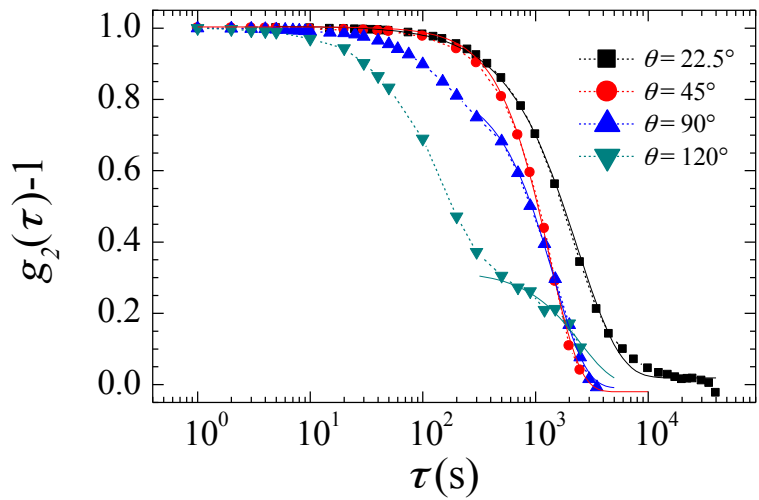

**Figure 2**. Intensity correlation function measured in a DLS experiment on a bare gel at 1.3% w/w of alginate. Data are collected sequentially at four scattering angles, as shown by the labels, at a temperature $T = 36°C$. The intensity correlation function are normalized by imposing $g_2(\tau = 1\ \text{s})-1 = 1$.

To investigate the microscopic origin of the relaxation behavior observed in rheology and DMA tests, we perform light scattering measurements on gels of the same composition. In Fig. 2 we show the dynamics of a bare gel, as probed by multi-speckle DLS performed at different angles. First, we note that a full decay is always visible, even at the lowest $q$. This indicates that the gel is remodeled on length scales at least as large as a few microns; this process is however very slow, since $g_2-1$ fully decays only on time scales of the order of thousands of seconds. Furthermore, a two-step decay is clearly observed at the largest $q$ vector, and is also visible for the data collected at $\theta = 90$ deg. Thus, the gel remodeling occurs via a complex relaxation process: as in the mechanical tests, we shall refer to the first relaxation process seen for the data at $\theta = 90$ or 120 deg as to the intermediate relaxation, while we will term the second decay the final relaxation (a faster, initial decay not visible here will be discussed in the following). Note that the amplitude of the intermediate decay decreases with decreasing scattering angle so that at the lowest $q$'s only the final decay is visible. This indicates that the intermediate decay involves motion of gel strands on a limited length scale, on the order of about 100 nm. Finally, we emphasize that the shape of the final relaxation is not a simple exponential decay. The lines in Fig. 2 show fits to the data using a stretched exponential function, $g_2-1 \propto \exp\left[-(\tau/\tau_f)^\beta\right]$, an expression often used in the context of slow glassy dynamics. While usually $\beta < 1$ for glassy systems, here we find surprisingly $\beta = 1.2 - 1.6$, depending on $q$. Values of $\beta$ larger than one indicate that the final relaxation is actually steeper than exponential, a behavior that has been termed compressed exponential and that has been reported for a number of jammed soft materials[29,35], including colloidal26[26] or biopolymer[28] gels. In gels, such dynamics have been ascribed to the relaxation of internal stress accumulated during the formation of a system-spanning network: internal stress contributes to bond breaking, leading to an intermittent series of dynamical events that slowly remodel the gel.[27]

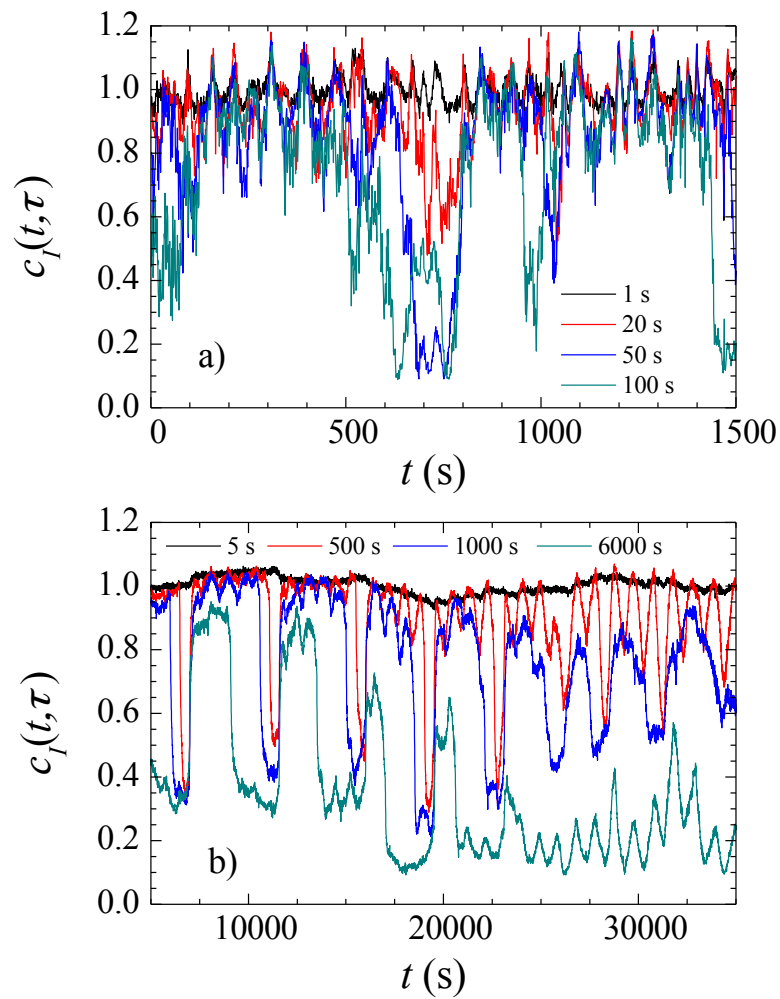

**Figure 3**. a): Temporal fluctuations of the time-resolved degree of correlation $c_I(t,\tau)$, for the same DLS experiment as in Fig. 2 and for $\theta = 120$ deg. The curves correspond to four values of the time delay $\tau$, as shown by the labels, all belonging to the range of the intermediate gel relaxation. For clarity, only a limited time window is shown. b): same as in a), but for $\theta = 22.5$ deg and for four $\tau$ values in the range of the final relaxation. In both cases, large fluctuations of the degree of correlation are visible, indicating temporally heterogeneous dynamics.

To test whether a similar scenario may also apply to our alginate gels, we inspect the temporal evolution of the time-resolved degree of correlation $c_I$. Figure 3 shows the TRC data from which the time-averaged correlation functions shown in Fig. 2 were obtained. Remarkably, our time-resolved measurements show that the relaxation processes probed by DLS are indeed highly discontinuous in time, both for the intermediate relaxation process (a), data for $\theta = 120$ deg) and for the final relaxation (b), data for $\theta = 22.5$ deg). The sudden drops of $c_I$ correspond to discontinuous, abrupt changes of the microscopic gel configuration, leading to a loss of correlation of the speckle pattern scattered by the sample. While our experiments are not performed in the space-resolved TRC configuration described in Ref. 25, where the size $\xi$ of the gel region affected by a rearrangement event can be directly measured, valuable information on $\xi$ can still be inferred by considering the amplitude of the drops of the degree of correlation shown in Fig. 3. As discussed in more details in 33, the very existence of large fluctuations of $c_I$ indicates that $\xi^3$ should be comparable to the size of the scattering volume, $V_s \approx 2\,\mathrm{mm}^3$. Indeed, if $\xi^3 \ll V_s$ a large number of rearrangement events would be required to remodel the whole scattering volume and decorrelate the scattered light. This would result in an efficient averaging over $V_s$ and thus to very small fluctuations of the TRC signal. A conservative estimate of $\xi$ may be obtained by assuming that a local event fully decorrelates the contribution to $c_I$ issued from the rearranged region, so that the degree of correlation drops to a value equal to the fraction of the scattering volume that is not rearranged. The data of Fig. 3 show then that each rearrangement event impacts more than half of the scattering volume, corresponding to $\xi > 1$ mm. Rearrangement events spanning several mm where indeed reported for the heterogeneous alginate gels of Secchi *et al.*[24] and for other gels and soft matter jammed systems.[25,28,34] For all these systems, such large values of $\xi$ are believed to stem from the elastic propagation of the strain field generated by a local rearrangement over large distances.[34]

Both the compressed exponential shape of the final decay of $g_2$-1 shown in Fig. 2 and the intermittent behavior highlighted in Fig. 3 are consistent with the scenario of a stress-assisted final relaxation of the gels. Previous work[29,35] has revealed another prominent feature of stress-relaxation-driven dynamics in soft matter, namely the anomalous $q$ dependence of its relaxation time, which typically scales as $q^{-1}$ for stress-driven dynamics, as opposed to $q^{-2}$ for a diffusive process. The data of Fig. 2 cannot be used to extract reliably the $q$ dependence of the relaxation time, since they were not taken simultaneously and, as we will show, the gel dynamics evolve with sample age. We therefore perform a separate experiment, where DLS data at three $q$ vectors are measured simultaneously. In order to better characterize the gel remodeling, we add to the sample a small amount of tracer particles, so that the scattering is dominated by the particle contribution and DLS data are essentially sensitive only to the motion of the tracers. This allows us to focus on the irreversible displacements traced by the particle motion, thus simplifying the data analysis. Additionally, the tracer particles enhance significantly the scattered signal. While the scattering is still sufficiently low for measurements to be analyzed in the DLS framework, a higher signal allows us to reduce the CCD exposure time and thus to access shorter delay times as compared to those for the bare gels.

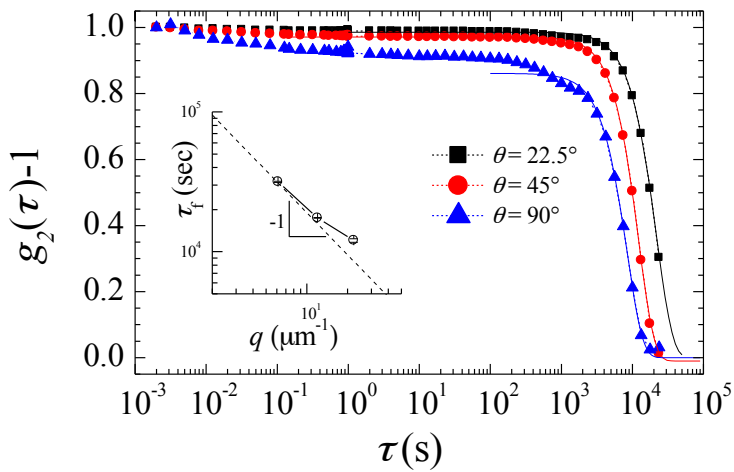

**Figure 4**. Main plot: Intensity correlation functions measured simultaneously by DLS at three scattering angles, for a gel at 1.3% w/w of alginate, loaded with a small amount ($\varphi_{DLS}$ = 0.0032%) of tracer particles of diameter $2a$ = 505 nm. Data were collected at $T$ = 36°C and at an age $t_w = 8 \times 10^4$ sec. The intensity correlation functions are normalized to unity at $\tau$ = 2 ms. Inset: $q$ dependence of the final relaxation time as obtained from a compressed exponential fit of the final decay of $g_2$-1 (lines in the main plot).

The main feature of Fig. 4 is similar to that observed in Fig. 2: a full decay of the intensity correlation function at all $q$ vectors, over several thousands of seconds. The final relaxation time is somehow larger than in Fig. 2, most likely because the gels of Fig. 4 were aged for a longer time. In analogy to Fig. 2, for $\theta = 90$ deg a two-step decay is observed, while at smaller $q$ vectors only the final relaxation is visible. As for the bare gels, the final decay can be fitted by a compressed exponential (solid lines in Fig. 4). Moreover, inspection of the time-resolved degree of correlation (data not shown) reveals that the dynamics of the particle-loaded gels are temporally heterogeneous, similarly to those for the bare gels. The overall agreement between the DLS data for bare and particle-loaded gels demonstrates that the small amount of tracer particles that were added does not change the gel dynamics and that the particles faithfully track the gel rearrangements. This last point indicates that the particles are large enough to be effectively trapped in the gel network and do not diffuse through its pores. This is consistent with estimations of the largest pore size in alginate gels, which are in the range of 120-150 nm.[36] Interestingly, an additional fast relaxation mode is visible at $\theta = 90$ deg on time scales of tens of ms, thanks to the shorter delay times accessible in the DLS measurements on particle-loaded gels. This fast mode is likely to be due to thermally activated, overdamped vibrations of the gel strands at fixed topology, *i.e.* on time scales shorter than those of rearrangement events associated with bond breaking. Similar dynamics have been reported in both experiments[37] and simulations[38] on colloidal gels, where it was also shown that the amplitude of the fast mode vanishes as $q \rightarrow 0$. This explains why the fast mode here can be observed only at the largest $q$ vector.

We fit the final relaxation of $g_2-1$ with a compressed exponential and show in the inset of Fig. 4 a double logarithmic plot of the $q$ dependence of $\tau_f$, the relaxation time of the final decay. The dashed line is the power law $\tau_f \propto q^{-1}$: the data at smaller $q$ appear to be consistent with this behavior, while they tend to level off at large $q$. By contrast, the data are clearly incompatible with the stronger length-scale dependence $\tau_f \propto q^{-2}$ predicted for diffusive-like dynamics. The temporally heterogeneous nature of the dynamics, the compressed exponential shape of the final relaxation of $g_2$-1, and the supradiffusive behavior of its relaxation time are strongly reminiscent of the slow dynamics of colloidal gels.[26,27] Thus, we propose that the physical mechanism responsible for the slow dynamics of the alginate gels is the same as for the colloidal gels: a slow network restructuring induced by bond-breaking and re-bonding events that are discontinuous in time and that are induced or at least enhanced by internal stress. A simple model of discontinuous dynamics for the colloidal gels has been discussed in Ref. 27: this model may in particular explain the leveling off of $\tau_f$ vs. $q$ seen in the inset of Fig. 4. The model assumes that the dynamics are due to discontinuous events that occur at an average rate $\Gamma$, each of them displacing the scatterers over a distance $\delta$. If the probed length scale, $2\pi q^{-1}$, is larger than $\delta$, many such events are required to decorrelate the scattered light. The smaller the scattering vector, the larger the number of events required for a full decorrelation, leading to a $q$-dependent relaxation time of $g_2$-1. By contrast, for scattering vectors large enough so that $2\pi q^{-1} < \delta$ one single event is sufficient to fully decorrelate the scattered intensity. In this case, the only relevant time scale for the decay of $g_2$-1 is the event rate and one expects $\tau_f \sim \Gamma^{-1}$, independently of $q$. Unfortunately, fully testing this scenario for our alginate gels would require extending the DLS measurements to $q$ vectors larger than those accessible by light scattering. Nevertheless, the data shown in the inset of Fig. 4 suggest that $\Gamma^{-1}$ should be on the order of several thousands of seconds, a value consistent with the typical time between events as revealed by the temporal spacing of the large drops of $c_I$ observed in Fig. 3b. The typical displacement associated with one single event would then be of the order of $\sim 2\pi/(30 \, \mu\mathrm{m}^{-1}) \sim 0.2 \, \mu\mathrm{m}$.

In spite of these analogies, the relaxation behavior of the alginate gels is more complex than that of colloidal gels, since two restructuring modes (the intermediate and final relaxation discussed above) are observed both in mechanical and DLS measurements, as opposed to one single slow mode for colloidal gels. In DLS, the intermediate mode that appears to be specific to the alginate gels can only be seen at the highest $q$ vectors, suggesting that it involves restructuring over length scales too small to be conveniently captured in single scattering experiments. We thus perform DWS measurements in the multiple scattering regime, which are sensitive to motion on shorter length scales. In order to obtain turbid samples, the alginate gels are loaded with particles larger than those used for DLS, so as to obtain strong multiple scattering conditions even at a modest particle concentration ($\varphi_{DLS} = 0.8\%$), because larger particles scatter light much more efficiently.

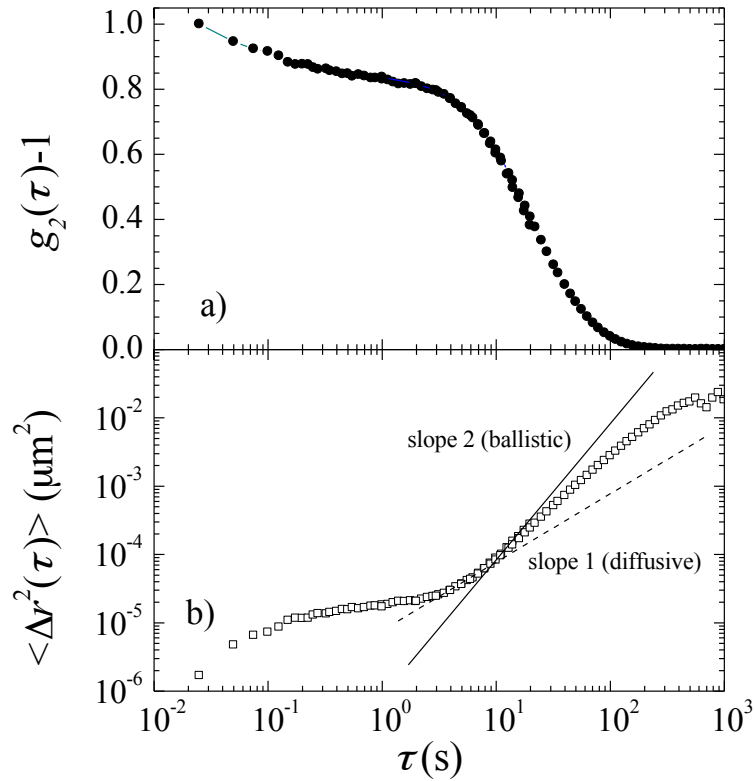

**Figure 5**. a): Intensity correlation function measured by DWS on a gel at 1.3% w/w of alginate, loaded with a small amount ($\varphi_{DWS}$ = 0.8%) of tracer particles of diameter $2a$ = 910 nm. Measurements are performed at $T$ = 38°C, after a waiting time large enough for sample aging to be negligible. Data are normalized to one at the smallest available delay. b): Mean squared displacement of the tracer particles obtained by inverting the data of a) using Eq. (1).

Figure 5a shows the DWS intensity correlation function measured in the backscattering geometry. As for the DLS measures reported in Fig. 4, we distinguish a fast decay on time scales of a few tens of ms, barely visible due to limitations in the CCD acquisition rate. This fast mode is again identified with the overdamped vibrations of the gel strands and is followed by a plateau and a full relaxation on a time scale of a few tens of seconds. Using the DWS formalism, Eq. (1), we obtain from $g_2$-1 the mean square displacement (msd) of the tracer particles as a function of $\tau$, shown in Fig. 4b. By comparing measurements for cell thicknesses of 5 mm and 10 mm, we confirm that our results are independent of cell thickness, consistently with Eq. (1) (data not shown). Up to $\tau \approx 2$ s, the msd exhibits the typical behavior for a soft solid: an initial growth that is nearly diffusive (msd ~ $\tau$) reflecting the thermally activated vibrations of the network, followed by a plateau, indicating that the probe is trapped in a solid-like structure. As proposed by Mason and Weitz[39], a generalized Stokes-Einstein relation may be used to estimate the elastic modulus $G'$ of the gel from the mean squared displacement at the plateau:

$$G' \approx k_B T / \left( \pi a \langle \Delta r^2 \rangle_{plateau} \right) \approx 190 \, \text{Pa} \tag{4}$$

where $k_B$ is Boltzmann's constant and where we have taken $\langle \Delta r^2 \rangle_{plateau} \approx 1.5 \times 10^{-5} \mu\text{m}^2$. A more refined calculation using the micro-rheology formalism proposed in[40] yields a similar value, $G'$ =

220 Pa at a frequency $\nu = 0.5$ Hz. This value is of the same order of magnitude of $G' = 730$ Pa as obtained from mechanical measurements; the discrepancy is likely due to the different protocol used for preparing the gels, in addition to the fact that micro-rheology is known to somehow underestimate elasticity, especially for heterogeneous materials.[41] Substantial differences between micro-rheology and mechanical rheology of gels have been reported when the probes are small enough to diffuse through the gel pores; here, the fair agreement between micro- and macro-rheology indicates that the particles are effectively trapped in the gel, as observed previously for the DLS data.

On time scales larger than a few seconds, the msd grows again, revealing a restructuring process that opens the cage in which the particles were initially trapped. This final regime spans roughly two decades, 10 sec $< \tau <$ 1000 sec and corresponds to a particle displacement up to ~100 nm. A comparison with the DLS data strongly suggests that this final increase of the msd, stemming from the final decay of the DWS correlation function shown in Fig. 5a), corresponds to the intermediate relaxation seen in the DLS data of Fig. 2 and 4. Further support to this identification comes from the observation that the dynamics probed in this regime are temporally heterogeneous, as revealed by the DWS time-resolved degree of correlation (data not shown), which exhibits a behavior similar to that shown in Fig. 3a for DLS. We emphasize that in this regime the msd grows with $\tau$ faster than linearly (see the dashed and solid lines in Fig. 5b) that show diffusive and ballistic behavior, respectively), indicating supradiffusive motion, close to ballistic. As discussed above, temporal heterogeneity and supradiffusive behavior are the hallmark of stress-enhanced network restructuring. Thus, our DWS experiments prove that the intermediate relaxation corresponds to a network remodeling process, similarly to the final relaxation.

Before proceeding further, it is interesting to compare in more detail our findings to those of Secchi et al.[24] who studied alginate solutions gelled via a different protocol, i.e. an ion diffusion mechanism. Analogies include the existence of a very slow relaxation mode associated with temporally heterogeneous dynamics and aging effects (for our gels, we shall discuss aging in the following). One obvious difference is the fact that only one relaxation mode was observed in Ref. 24, to be contrasted with the three modes discussed here. However, the setup of Secchi and coworkers lacked the temporal resolution required to measure the fast and intermediate relaxation processes reported here, so that one cannot rule out their existence for the gels of Ref. 24. More importantly, the shape of the final relaxation was found to be a simple exponential by Secchi et al, as opposed to a compressed exponential for our gels and for most soft systems where internal stress is likely to be at the origin of the slow dynamics. Secchi and coworkers propose that the simple exponential decay observed in their experiments corresponds to one of the regimes predicted in earlier theoretical work on the dynamics of colloidal gels by Bouchaud and Pitard.[42] The analysis of Ref. 42 is particularly simple if the experimental scattering vector satisfies the inequality $q \ll E\Theta/(\eta\xi)$ (see Sec. 5 therein), where E is the elastic modulus of the gel, $\xi$ its mesh size, $\eta$ the solvent viscosity and $\Theta$ the duration of a rearrangement event (a "micro-collapse" in the language of Ref. 42). This inequality is certainly satisfied for the gels of Ref. 24, for which the authors report $E \approx 200 kPa$, $\xi \approx 10$ nm, $\eta \approx 10^{-3}$ Pa s, and $\Theta \approx 2000$ s (as obtained from the duration of the drops of cI shown in the inset of Fig. 10), leading to $q \approx 2 \times 10^7$ m$^{-1}$ $\ll E\Theta/(\eta\xi) \approx 4 \times 10^{19}$ m$^{-1}$. According to Ref. 42, in this regime a compressed (resp., simple) exponential relaxation is expected for $\tau \ll \Theta$ (resp., $\tau \gg \Theta$). The data presented in Ref. 24 exhibit a simple exponential relaxation in the regime $\tau < \Theta$, in contrast with the predictions of the model by Bouchaud and Pitard. Whatever the origin of the shape of $g_2$-1 for the gels of Secchi et al., the discrepancy between the results of Ref. 24 and those reported here highlight the sensitivity of the microscopic dynamics of alginate gels on composition and gelation protocol and calls for new, more refined theoretical approaches.

We now come back to the discussion of our light scattering results. Collectively, the DWS and DLS data indicate that our alginate gels undergo a complex restructuring process involving motion of gel strands on length scales from about 10 nm up to at least a few $\mu$m. This process involves two distinct time scales (the intermediate and final relaxation modes), in addition to over-damped vibrations of the network at short times. Both restructuring modes exhibit supra-diffusive motion and dynamical heterogeneity, as in other network-forming soft materials where internal stress relaxation has been invoked as the microscopic mechanism responsible for these slow dynamics. Internal stress thus appears as a natural candidate to explain the dynamics of our gels. However, we emphasize that DLS measurements on alginate solutions prior to gelation, where internal stress is unlikely to be present, have evidenced a relaxation mode ascribed to thermally activated bond breaking.[23] In gelled samples, both thermal energy and internal stress may then contribute to bond breaking. In order to better understand the relevance of either mechanism, we investigate the temperature and age dependence of the dynamics of our gels.

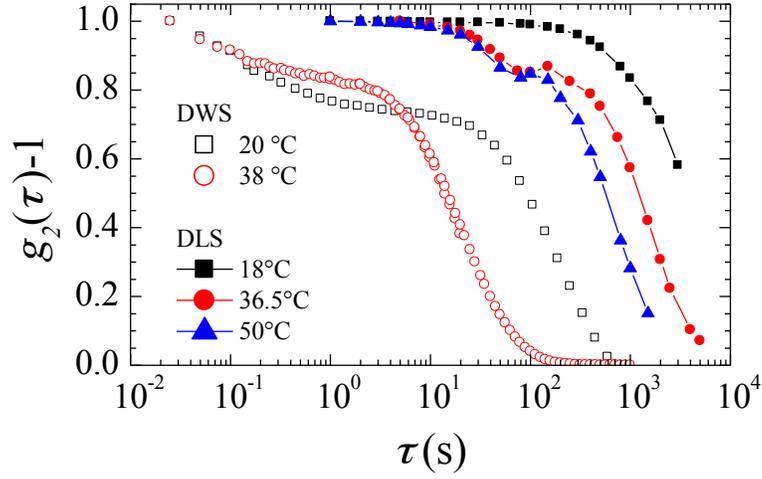

**Figure 6**. Temperature dependence of the intensity correlation function measured by DLS for bare gels at $\theta = 90$ deg (solid symbols) and by DWS on gels loaded with particles with $2a = 910$ nm at $\varphi_{DWS} = 0.8\%$ (open symbols). All correlation functions reported here were measured for sample ages long enough for aging effects to be negligible. Data are normalized to one for the shortest available delay.

Figure 6 shows that both the intermediate and the final relaxation processes become slower upon decreasing temperature, a general behavior observed also for the final relaxation of the Young modulus shown in Fig. 1. We first analyze the DLS data for the final relaxation, for which measurements at three temperatures are available. We expect the final relaxation time $\tau_f$ to be proportional to the life time of the longest living bonds, $\tau_l$. Assuming that both thermal activation and internal stress are responsible for bond breaking, one has:

$$\tau_f \propto \tau_l \propto \exp[(U_l - F_l b_l)/k_B T] \qquad (5)$$

where the subscript $l$ denotes long-lived bonds and where $U_l > 0$ is the bond energy, $b_l$ the bond length, $k_B T$ the thermal energy and $F_l$ the force acting on the bond due to internal stress. An Arrhenius plot of the DLS final relaxation time versus $T$ yields $U_l - F_l b_l \approx 6.5 \times 10^{-20}$ J. This value is comparable to but significantly smaller than $U_{eb} = 7 \times 10^{-19}$ J, the bond energy for one single egg-box link, as calculated in Ref. 43. Since the bonds responsible for the slowest relaxation mode are the longest lived, they must involve several egg-box units or even the cross linking of more than two chains in a bundle. Such structures are more stable than a single egg-box link; therefore one expects an effective bond energy $U_l >> U_{eb}$. Since in our case $U_l - F_l b_l < U_{eb}$, we conclude that the force acting on a bond must be non-zero, i.e. that internal stress contributes significantly to the slowest relaxation mode of the gels. A similar analysis on the intermediate relaxation yields $U_s - F_s b_s \approx 1.4 \times 10^{-19}$ J, where the subscript $s$ refers to the shorter-lived bonds whose breaking is responsible for the intermediate mode. Again, we find $U_s - F_s b_s < U_{eb}$, implying that internal stress plays a role also in the intermediate relaxation mode. Although one should be cautious in analyzing the intermediate mode given that only two temperatures are available for the DWS data, we note that $U_s - F_s b_s > U_l - F_l b_l$. Since the bond energy is likely to be lower for the shorter-lived bonds, this inequality suggests that $F_s b_s < F_l b_l$. The bond length should be the same for both short-lived and long-lived bonds, because the molecular structure of the links is the same -G monomers ionically cross-linked by the $Ca^{2+}$ ions-, the only difference being the number of unit links participating in the bond. Thus, the comparison between the $T$ dependence of the intermediate and final relaxation times suggests that the force acting on the long-lived bonds is larger. In other words, the long-lived bonds from the scaffold that sustains most of the internal stress of the gel.

The picture emerging from the data discussed so far points to a double relaxation process driven by internal stress and thermal energy. Within this picture, the sample is endowed with a given amount of internal stress unavoidably generated during gelation, since gels form in a microscopically heterogeneous configuration. The links resulting from ionic bonds are expected to be reversible in time[44]; thus, bonds break and reform with time, thereby locally relaxing internal stress. One therefore expects the dynamics to slow down with sample age, because the driving force, *viz* the internal stress, would decrease progressively. To test this scenario, we report DLS and DWS measurement conducted on the same gel sample under the same experimental conditions, but at different ages. Preliminary tests show that the gel dynamics is very sensitive to mechanical and thermal perturbations: upon a change of $T$ or simply after displacing the sample to load it in the light scattering apparatus, the gel dynamics is initially relatively fast and slows down progressively with time, $t_w$. A similar "rejuvenation" effect upon mechanical perturbation has been reported for glassy soft matter[45]; for our gels, it is observable even after letting the gel rest unperturbed for several months. Accordingly, in the following, we define $t_w=0$ as the loading time before a series of measurements.

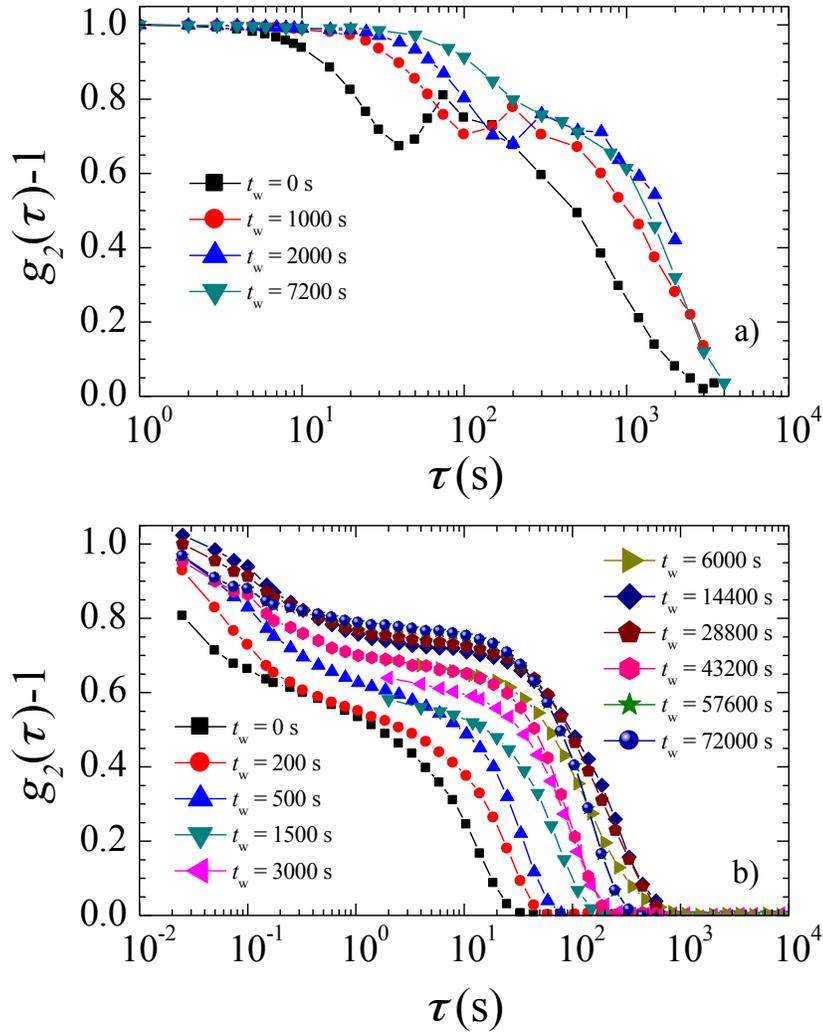

**Figure 7**. Aging behavior of the microscopic dynamics of a gel at 1.3% w/w of alginate as measured by a): DLS (bare gel, $\theta$ = 90 deg, $T$=36 °C) and b) DWS (gel loaded with particles with $2a$ = 910 nm at $\varphi_{DWS}$ = 0.8%, $T$=20°C). In both a) and b), $g_2$-1 for the smallest age has been normalized to one at the shortest available delay; all subsequent correlation functions are normalized using the same scaling factor. See the text for the definition of $t_w$=0.

Figure 7 shows the aging behavior as probed by DLS (a), bare gel, $\theta = 90$ deg) and DWS (b). The DLS data show that both the final and the intermediate relaxation become slower with time, a behavior characteristic of out-of-equilibrium systems. The final relaxation appears to reach a nearly stationary regime after a few thousand seconds; by contrast, the intermediate decay continues to slow down throughout the whole duration of the DLS experiment reported here (7200 s). This is consistent with the DWS data of Fig. 7b), where the intermediate relaxation process ages over a longer time span, until about 14000 s. Interestingly, a somehow erratic behavior is observed at later ages, where the dynamics may temporarily accelerate (compare e.g. the data for $t_w = 43200$ s to those at $t_w = 28800$ s). A similar behavior has been reported for colloidal gels.[25,46] It is likely due to the highly heterogeneous nature of the dynamics, which results in large fluctuations of $g_2-1$ as obtained by averaging $c_I$ over short time windows. Such effects are clearly visible in a nearly stationary regime, as seen in Fig. 7b for $t_w >\sim 10000$ s, while they are masked by the strong aging trend at earlier times.

Both mechanical and light scattering measurements show that our alginate gels exhibit a complex slow relaxation involving two distinct time scales, in addition to fast dynamics that we ascribe to network vibrations at fixed topology. The intermediate mode seen in rheology has a time scale of 10-100 s, in excellent agreement with the time scale of the sharp increase of the msd observed in the DWS measurements of Fig. 5b): it is therefore natural to identify these two modes. We mentioned that in rheology tests this intermediate mode accounts for the relaxation of only a modest fraction of the applied stress. This is consistent with the notion that the intermediate relaxation involves motion of gel strands only on a limited length scale, of the order of 100 nm, as inferred by DWS and DLS measurements.

The time scale of the final relaxation probed by DLS depends on the scattering angle. However, by invoking arguments originally put forward to model the intermittent dynamics of colloidal gels, we have proposed that this mode stems from rearrangement events occurring on a time scale of several thousands of seconds. This is fully consistent with the time scale of the final relaxation observed in rheology (see Fig. 1), whose onset occurs at $\tau \geq 1000$ s. We thus identify the final relaxation mode observed in rheology with that seen in DLS.

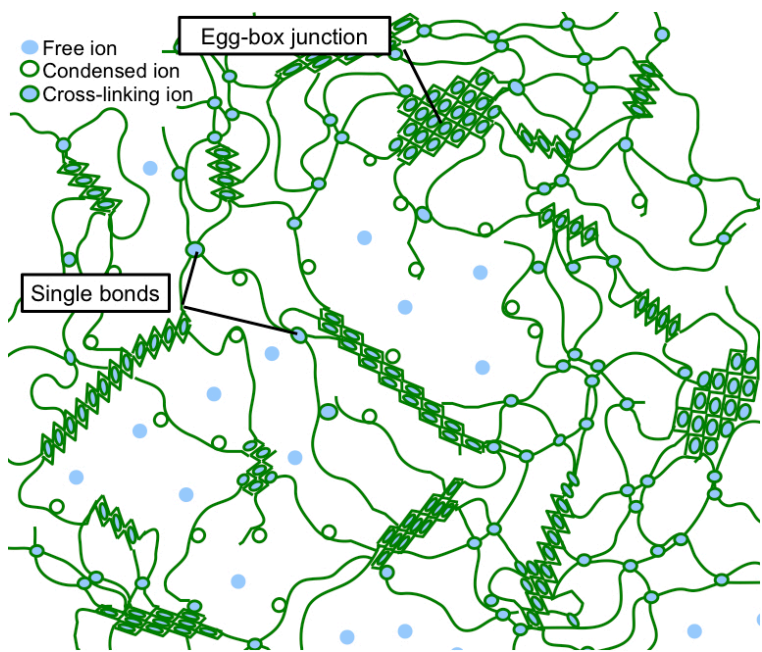

**Figure 8**. Schematic representation of the alginate gel, showing the hierarchical structure of cross-linking elements in the network, some of which involve short-lived bonds due to single ions or egg-box units, while longer-lived bonds are due to multi-unit egg-box structures, resulting in rod-like junctions and bundles of several chains.

Our microscopic measurements shed light on the origin of these modes. The gel dynamics can be ascribed to the existence of internal stress generated upon gelation, together with the fact that this stress can be progressively relaxed via bond-breaking and re-bonding events, since the crosslinks are not permanent. Both thermal energy and stress are thus responsible for the gel remodeling. The fact that the relaxation occurs through two steps suggests a hierarchical structure of bonds, some of which have a life time much longer than that of the others, because of their structure (*e.g.* several egg-box units and bundles as opposed to single cross-linking ions or single egg-box units), as well as due to the fact that internal stress is not homogeneous throughout the sample. Based on the discussion reported above, we sketch in Fig. 8 a schematic representation of the structure of the alginate gel. Densely cross-linked zones coexist with poorly cross-linked regions, i.e. water pools with dissolved ions. Moreover, the cross-linking ions can either participate to the formation of single cross-linking points or be involved in more complex, longer lived linking structures involving several egg-box units. This hierarchical structure endows the alginate gels with the complex relaxation behavior highlighted in this work.

One important conclusion of our measurements is the notion that internal stress plays a key role in the gel dynamics, even when *in situ* gelation methods are used to insure a macroscopically homogeneous material. This suggests that applying an external stress may allow one to control the gel restructuring and tune its dynamical properties. Measurements that couple simultaneously rheology and light scattering will be required to explore this intriguing possibility.

**Corresponding Authors**

Correspondence should be addressed to:

larobina@unina.it

luca.cipelletti@univ-montp2.fr

ACKNOWLEDGMENTS

Funding from CNES (France) and Ministero dell'Istruzione dell'Università e della Ricerca MIUR (project PRIN 2010-2011 Prot. 20109PLMH2_005) is gratefully acknowledged. We thank F. Docimo and R. Valentin for helping in the DMA and light scattering measurements, respectively.